# RADIATION HARDNESS OF GaAs:Cr AND Si SENSORS IRRADIATED BY ELECTRON BEAM

**U. Kruchonak[a,1], S. Abou El-Azm[a], K. Afanaciev[b], G. Chelkov[a], M. Demichev[a], M. Gostkin[a], A. Guskov[a], E. Firu[c], V. Kobets[a], A. Leyva[a,d], A. Nozdrin[a], S. Porokhovoy[a], A. Sheremetyeva[a], P. Smolyanskiy[a], A. Torres[e], A. Tyazhev[f], O. Tolbanov[f], N. Zamyatin[a], A. Zarubin[f] and A. Zhemchugov[a]**

[a] *JINR, Dubna, Russia,* [b] *INP, Minsk, Belarus,* [c] *ISS, Bucharest, Romania ,* [d] *CEADEN, Havana, Cuba*
[e]*InSTEC, Havana, Cuba,* [f]*TSU, Tomsk, Russia*

[1] Corresponding author *E-mail*: Uladzimir.Kruchonak@cern.ch

**Abstract**

The interest in using the radiation detectors based on high resistive chromium-compensated GaAs (GaAs:Cr) in high energy physics and others applied fields has been growing steadily due to its numerous advantages over others classical materials. High radiation hardness at room temperature stands out and needs to be systematically investigated. In this paper an experimental study of the effect of 20.9 MeV electrons generated by the LINAC-200 accelerator on some properties of GaAs:Cr based sensors is presented. In parallel, Si sensors were irradiated at the same conditions, measured and analyzed in order to perform a comparative study. The target sensors were irradiated with the dose up to 1.5 MGy. The current-voltage characteristics, resistivity, charge collection efficiency and their dependences on the bias voltage and temperature were measured at different absorbed doses. An analysis of the possible microscopic mechanisms leading to the observed effects in GaAs:Cr sensors is presented in the article.

## 1. Introduction.

Silicon detectors generally satisfy the requirements of the modern physics experiments, however, the experimental physics develops in the direction of increasing radiation loads so the requirements for the detector materials radiation hardness are increasing. Furthermore, for some experimental applications it is also important to have detectors with high atomic number. One example of such application is high energy X-rays registration. Another example is the operation of a semiconducting detector itself as an active target. Classical Si semiconductor detectors do not satisfy these conditions [1, 2, 3].

Other materials are used to solve these problems, such as diamond or compound semiconductors: cadmium telluride (CdTe), mercury iodide ($HgI_2$), thallium bromide (TlBr), Silicon Carbide (SiC) and gallium arsenide (GaAs) [2, 20].

Chromium-compensated gallium arsenide (GaAs:Cr) material was employed for the similar purpose. The advantages of this material are its high radiation resistance and high Z. Additionally, recent improvements in the manufacturing technology allowed production of sufficiently large thicknesses and areas of this material in an economically feasible way [4].

The study of the radiation hardness of this material is obviously essential for future applications. Some research on this topic with different types of radiation was already performed in the previous studies [5, 6]. Specifically, behavior of GaAs material compensated with chromium under irradiation was investigated in [7, 8]. High energy electron irradiation effects on electrical and spectroscopic properties of the GaAs detector were studied in [9]. It reported the increase of leakage current and the decrease of detector's charge collection efficiency (CCE) for GaAs: Cr sensors irradiated by 8.5-10 MeV electrons up to an absorbed dose of 1.5 MGy.

This paper describes the behavior of GaAs:Cr and silicon sensors irradiated with 20.9 MeV electron beam up to an absorbed dose of 1.5 MGy. The current-voltage characteristics (I-V) and the CCE were measured as a function of the absorbed dose. The measurements results are discussed and some possible explanations of the observed effects are presented.

## 2. The GaAs:Cr and Si sensors.

In this work, we studied semi-insulating GaAs:Cr sensors made of n-GaAs material using the precision chromium doping technique. The impurity concentration in GaAs: Cr does not exceed $2\times10^{17}$ $cm^{-3}$. The properties of this material determine the decisive participation of electrons in the charge collection due to the low value of mobility×lifetime product for holes $(\mu\times\tau)_p$. Thus CCE is close to 50% for unirradiated sensors at 100% electron collection.[7, 10]. These sensors were produced at Tomsk State University.

As a reference for the radiation hardness study and for the sake of direct comparison with GaAs:Cr, two n-type silicon pad sensors were studied. One of them was n-type Si made from FZ-Si-n (Wacker), with initial impurity concentration for carbon and oxygen less than $2\times10^{16}$ $cm^{-3}$, orientation <111> and $2 < \rho < 4$ kΩ×cm (RIMST, Zelenograd, Russia). The other was n-type Si made by Hamamatsu Photonics (HPK), orientation <100> and $1.25 < \rho < 3.25$ kΩ×cm, used as test structures during silicon strip sensor production for Fermi Gamma-ray Space Telescope [11].

Atomic displacement energy $E_d$ is 11-22 eV for Si and 9-9.4 eV for GaAs, when the minimum energy of electrons needed to create a displacement defect in crystals is 115-330 keV for Si and 230-260 keV for GaAs [19]. Some of the sensors were originally mounted on PCBs, and for others (kept as bare sensors) plastic holders were made. The characteristics of studied sensors are listed in Table 1 and their examples are presented in Figure 1.

**Table 1:** Si and GaAs:Cr sensors selected for the investigation.

| Sensor type | Sample № | Holder | Size ($mm^3$) | Sensitive area ($mm^2$) | Donor concentration ($cm^{-3}$) |
|---|---|---|---|---|---|
| GaAs:Cr (1) | N1, N4, N5, N6, N22 | Plastic | 5×5×0.3 | 5×5 | ~ $10^{17}$ |
| GaAs:Cr (2) | N7, N19, N21, N23 | PCB | 5×5×0.3 | 4.5×4.5 | ~ $10^{17}$ |
| Si (1) (RIMST) | N1, N3, N5 | Plastic | 5×5×0.25 | 3.6×3.6 | (0.8-1.6)×$10^{11}$ |
| Si (2) ( HPK) | 6884, 6886, 6888 | PCB | 5×5×0.4 | 3.5×3.5 | (1.3-3.5)×$10^{11}$ |

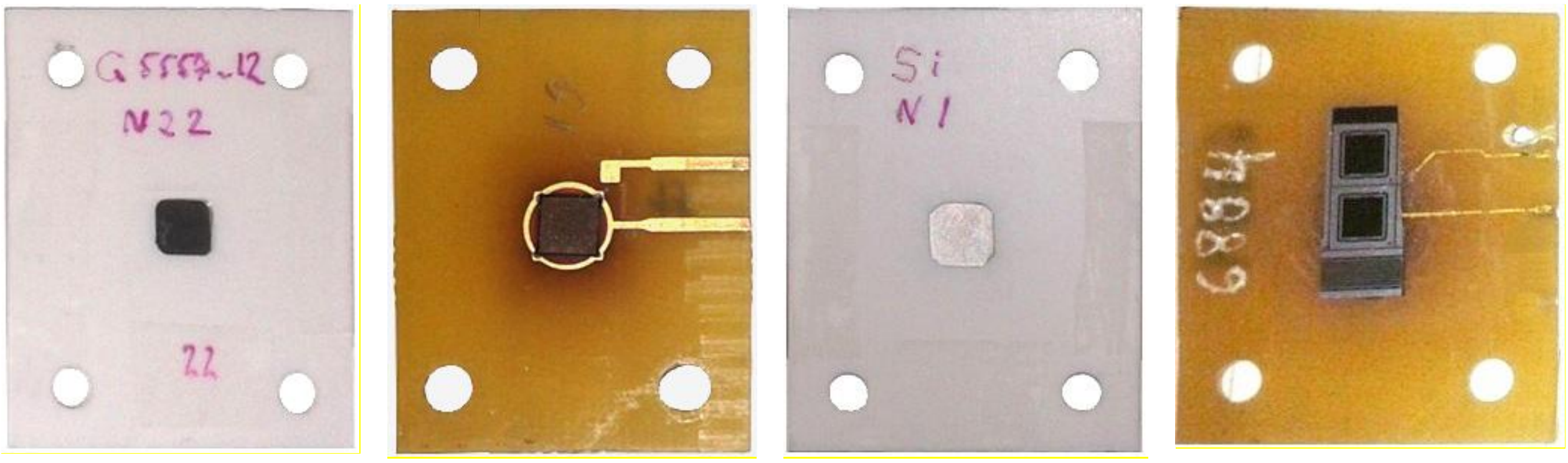

**Figure 1:** *Examples of each studied sensor type from set I. From left to right: GaAs:Cr (1) in plastic holder, GaAs:Cr (2) on PCB, Si (1) in plastic holder and Si (2) on PCB.*

## 3. Irradiation setup and absorbed dose control.

The electron irradiation was performed at the Joint Institute for Nuclear Research (Dubna, Russia) on theLINAC-200 accelerator using the 20.9 MeV beam channel. The electron beam energy was measured using the activation analysis method [12]. The electron beam was bunched with the bunch duration 2 μs, current up to 10 mA and the bunch frequency from 1 to 10 Hz.

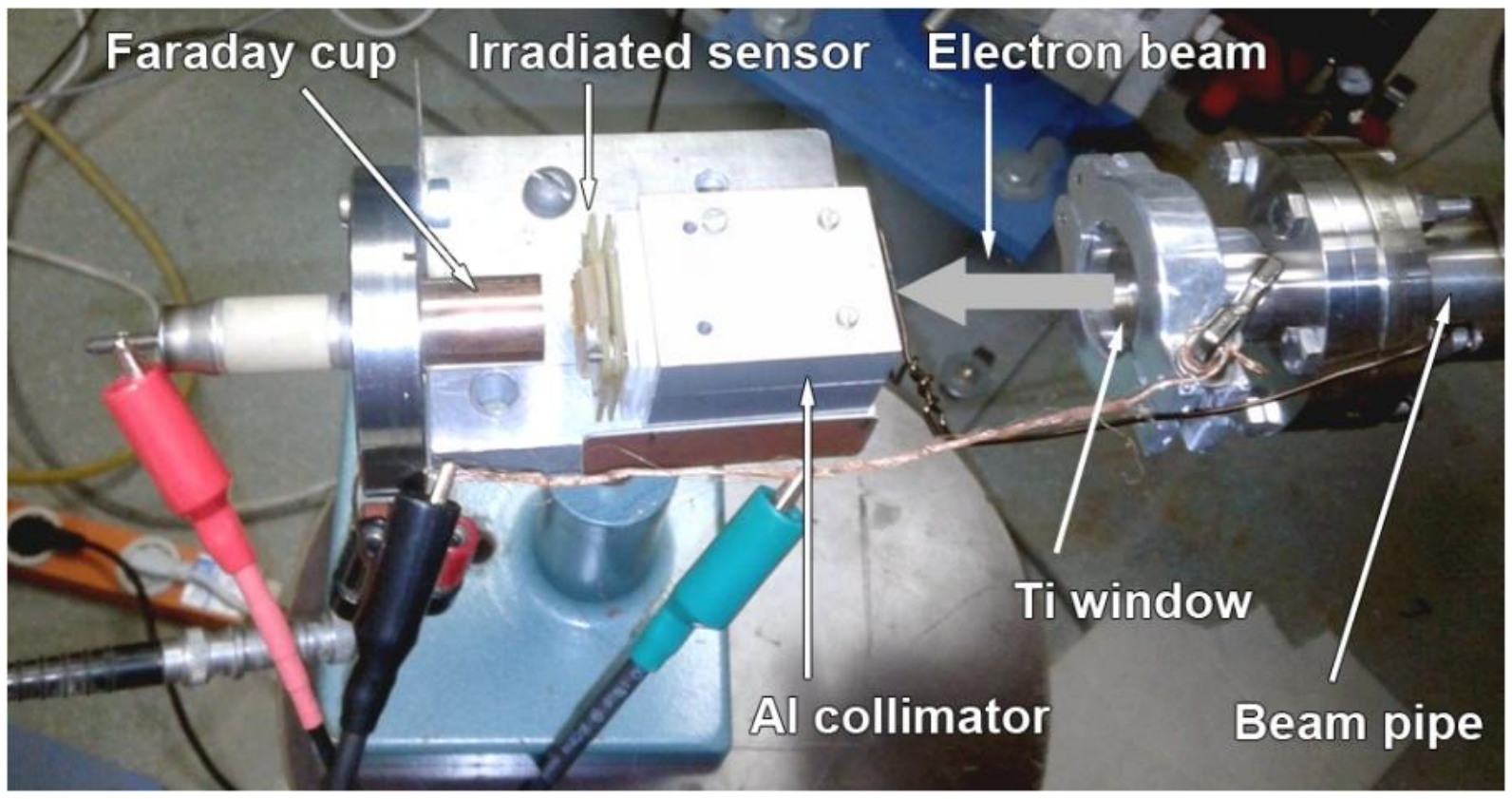

**Figure 2:** *Irradiation setup. The electron beam from the accelerator is shaped by the collimator, passes through the sensor with radiochromic film and finally reaches the copper Faraday cup.*

The irradiation setup is shown in Figure 2. The electron beam exits the beam pipe through a titanium window, traverses a distance of 60 mm through the air and then enters a 50 mm long aluminum collimator. The collimator shapes the beam to the 5x5 mm$^2$ square with approximately flat intensity profile. The size of the collimated beam roughly corresponds to the sensor size to provide uniform exposure of the sensor area.

The collimated electron beam hits the target sensor, passes through the radiochromic film and is finally absorbed by the Faraday cup.

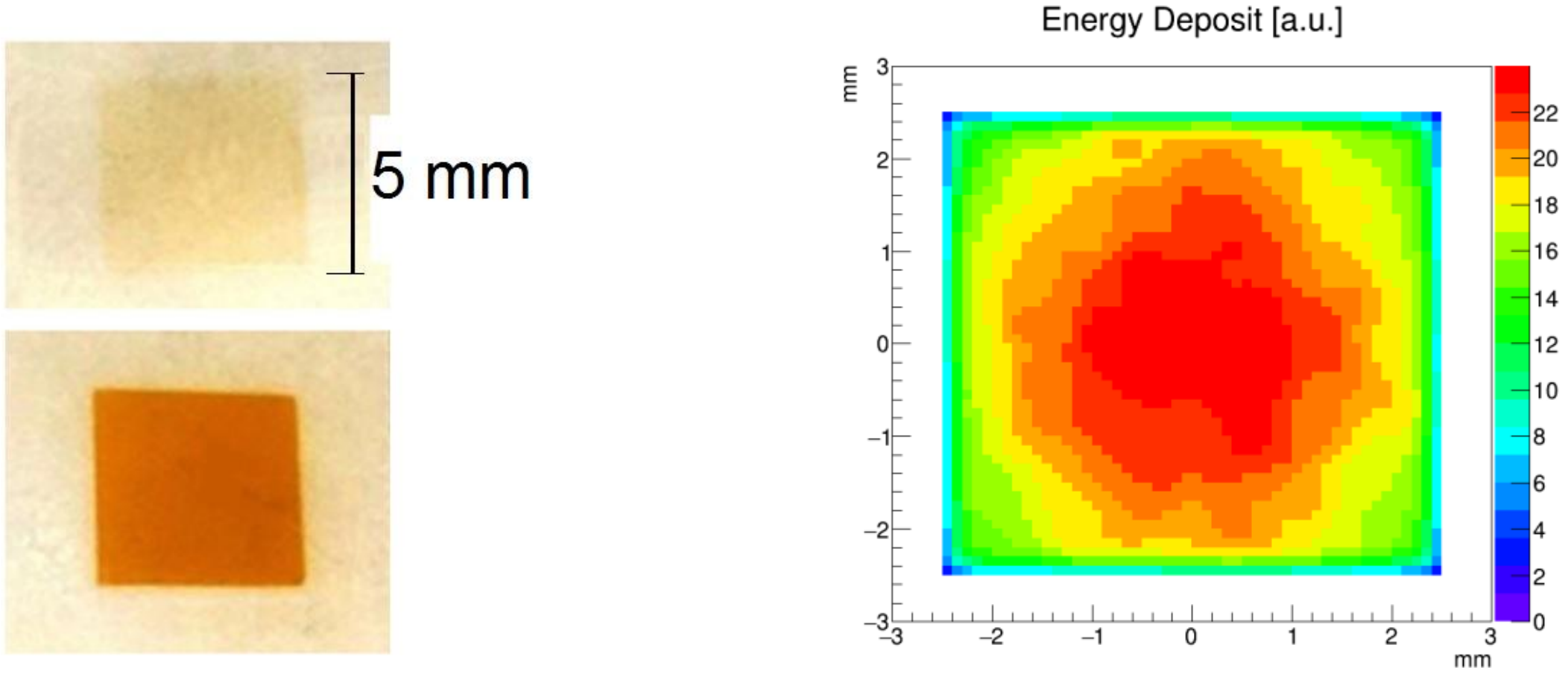

**Figure 3:** (color online) *Images of two irradiated radiochromic films (left) after the exposition to the electron fluence 1.9×10$^{14}$cm$^{-2}$ (above) and 8.2×10$^{14}$cm$^{-2}$ (below), corresponding to doses 43 and 171 kGy in GaAs:Cr respectively, and the GEANT4 simulation of the deposited energy distribution over 5x5x0.3 mm3 GaAs:Cr sensor volume ( right).*

The measurement of the charge in Faraday cup allows to estimate the electron flux and to calculate the absorbed dose. The integral electron flux varied in the range 0.5 - 5 x 10$^{14}$ e$^-$/hour, which corresponds to the absorbed dose rate from 50 to 500 kGy/hour. The radiochromic film provides additional estimate of the absorbed dose and control over the electron beam profile uniformity. Figure 3 presents two examples of the observed changes in the radiochromic film after irradiation and corresponding energy distribution over the sensor area as simulated in GEANT4 [13]. The difference in the dose between the center of the sensor and its edges is estimated to be about 20%. Three sensor sets were prepared for irradiation, each including two bare (type 1) and one PCB mounted (type 2) GaAs:Cr, one Si (type1) and one Si (type2) sensors. All sensors from the set I were irradiated simultaneously with steps of 25 kGy up to 100 kGy absorbed dose. Simultaneous irradiation was considered unsuitable due to beam scattering during passage and sensors from sets II and III were irradiated one by one or in pairs with steps from 50 to 200 kGy, up to absorbed dose of 0.5 MGy for the set II and up to 1.5 MGy for the set III.

The electron transport in the irradiation setup was simulated in GEANT4 in order to calibrate the Faraday cup charge measurement to the absorbed dose units. The final simulation results are presented in Table 2. Statistical error of simulation does not exceed 0.2%. Systematic error of simulation was estimated by changing low-energy electromagnetic model of GEANT4 simulation from Livermore to Penelope. The effect is about 3%. The simulation, however, predicts non-uniformity of energy deposit in the sensor on the level of 20% (Figure 3, right) while darkening of radiochromic film exposed area (Figure 3, left) is quite uniform. The effect of non-uniformity is possibly due to simple pencil beam model with fixed position which might differ from actual beam conditions. It is considered to be the main contribution to the systematic error.

**Table 2:** Results of GEANT4 simulation for dose calibration.

| Configuration | Holder | Thickness (μm) | Absorbed dose to charge in Faraday cup (kGy/μC) | |
|---|---|---|---|---|
| | | | 1-st sensor | 2-nd sensor |
| GaAs:Cr (1) | Plastic | 300 | 4.99 | - |
| GaAs:Cr (2) | PCB | 300 | 4.99 | - |
| Si type1 (3) | Plastic | 250 | 5.61 | - |
| Si type2 (4) | PCB | 400 | 5.68 | - |
| (1) + (1) | Plastic | 300/300 | 5.14 | 4.87 |
| (2) + (4) | PCB | 300/400 | 5.19 | 5.51 |
| | | | | |

## 4. I-V and charge collection measurements setup.

After each beam exposure the CCE for at least two bias voltages and the I-V characteristics were measured. The diagram of the CCE measurement setup is shown in Figure 4. The collimated electrons from the $Sr^{90}$ ß-source pass across the biased sensor producing a certain number of electron-hole pairs by ionization. These carriers are collected in the applied electric field and the resulting signal is read out by charge sensitive amplifier Amptek A250. The signal from the amplifier is finally digitized by DRS4 Evaluation Board [14]. The data is read out via USB to a computer for processing.

Data acquisition in the DRS4 Evaluation Board is triggered by coincidence of signals from two scintillation counters with different thicknesses. This arrangement allows digitizing only the signals produced by electrons that cross the sensor and have the energy in the range from 1 to 2.2 MeV. The average (and the most probable value, MPV) number of e-h pairs per 1 μm generated in the studied sensors by such electrons was simulated by GEANT4 and is estimated at 254 (MPV=154) for GaAs: Cr and 102 (MPV=77) for Si [13]. Such electrons can be considered as minimum ionizing particles (MIPs) with a well-known energy deposition in the material. The pedestal spectrum was collected separately, trigger was started from the generator under the same conditions as when collecting the MIP spectrum (Figures 4, 5).

Offline integration of a digitized signal waveform above an automatically determined baseline level was carried out to obtain the integrated charge values. We assume that this integral is proportional to the collected charge in the sensor or CCE. Thus, the ratio of the integral before and after irradiation is equal to the ratio CCE / $CCE_0$.

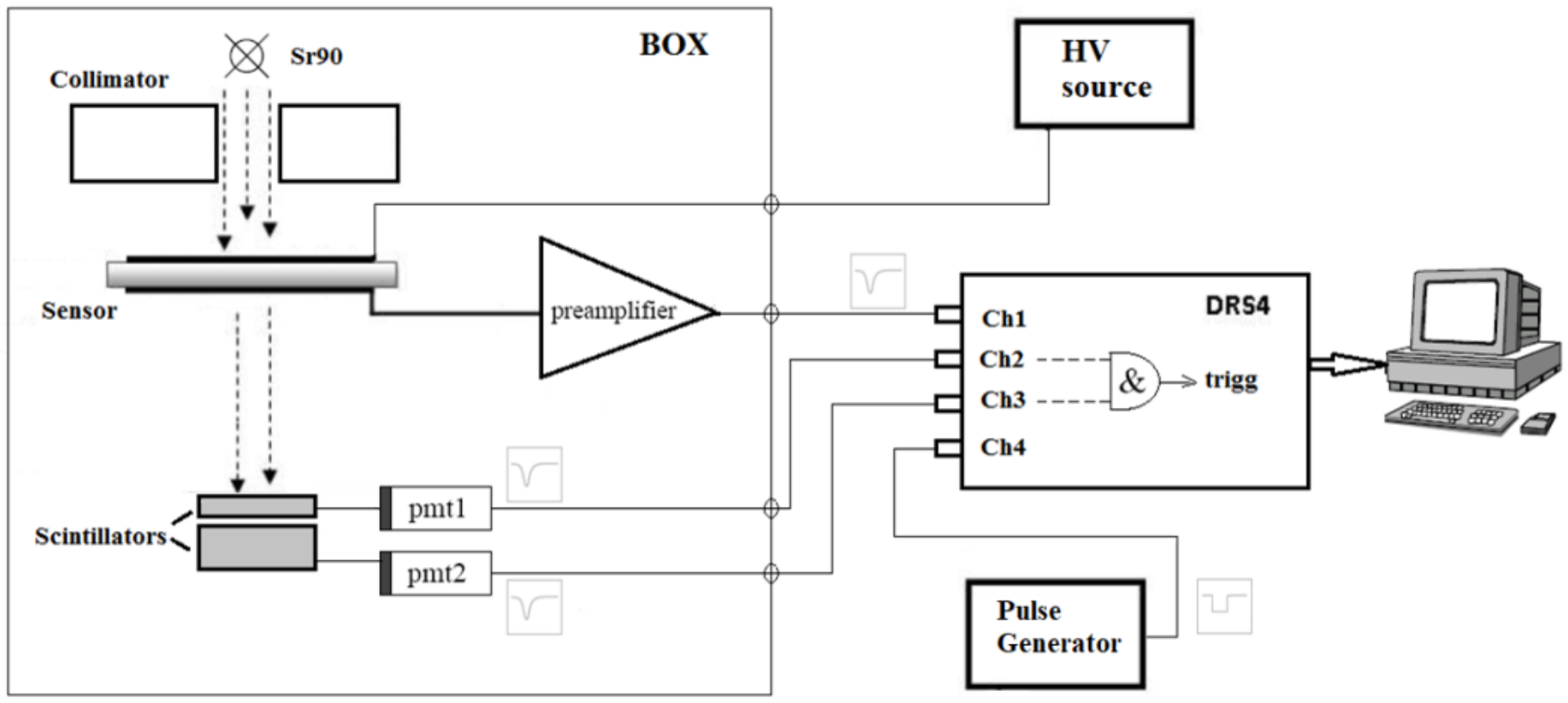

**Figure 4:** *Block diagram of CCE measurement setup.*

Figure 5 shows the measurement setup for CCE and I-V measurements. The setup is contained in a shielded and grounded light-tight metal box. The good collimation of the $Sr^{90}$ source and trigger scintillator setup allows to suppress pedestal counts almost completely. As a result, it is possible to measure very low CCE values (about 1%). A programmable voltage source coupled to the remotely controlled ammeter is used for the measurement of the I-V characteristics. The measurement setup is temperature stabilized. The amplifier integration time is 50 ns and intrinsic noise of the measuring system (FWHM) is about 2.5 $ke^-$ with the trigger activated and about 2 $ke^-$ with self-trigger. This setup allows to measure sensor parameters both at room temperature and preset low temperature.

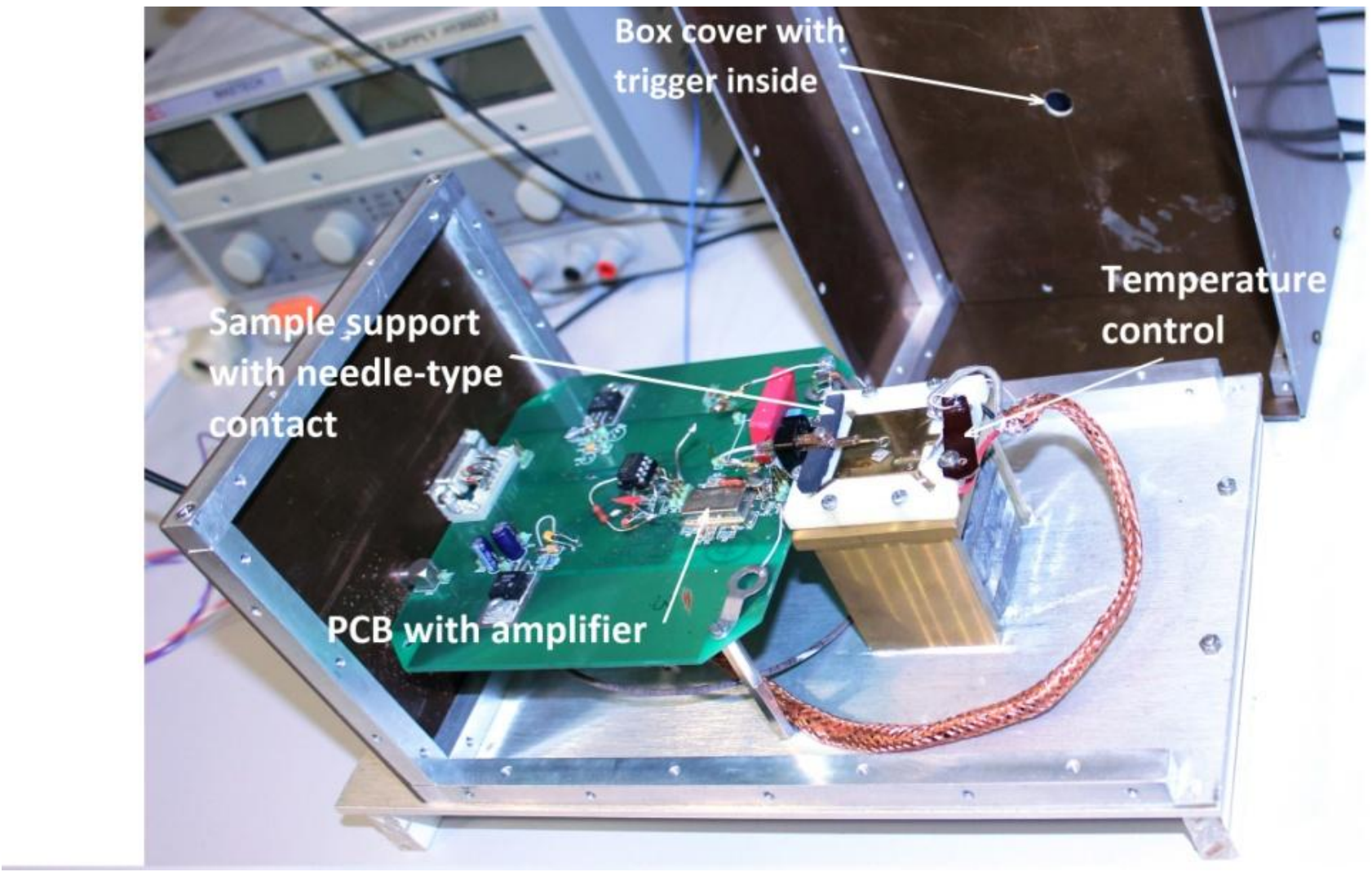

**Figure 5:** *A photography of the setup for CCE and I-V measurements.*

## 5. Results and discussion.

### 5.1. *I-V and resistivity measurements.*

Figures 6 and 7 present the I-V characteristics of GaAs:Cr and Si sensors and the effective resistivity of GaAs:Cr irradiated with different doses. Since there is no p-n junction in GaAs:Cr sensors, its effective resistance was calculated as $R_{eff} = U_{bias} / I$, where $U_{bias}$ = -200 V is the operating voltage for GaAs: Cr sensors, this voltage was mainly used for CCE measurements.

Figure 6 (left) shows that at the absorbed dose of 1.35 MGy for the GaAs:Cr N6 the dark current increases approximately 7-fold at a bias voltage of -200 V. Figure 6 shows that the effective resistivity of different samples decreased with the absorbed dose.

Some authors, such as [6, 15], argue that this behavior is fundamentally due to the introduction of displacement defects in the material lattice due to radiation damage. Such defects leading to formation of deep levels in the band gap can act as recombination-generation centers. Moreover, the radiation induced defects in the middle of the band gap are efficient electron-hole pair generation centers and thus are also responsible for the leakage current increase.

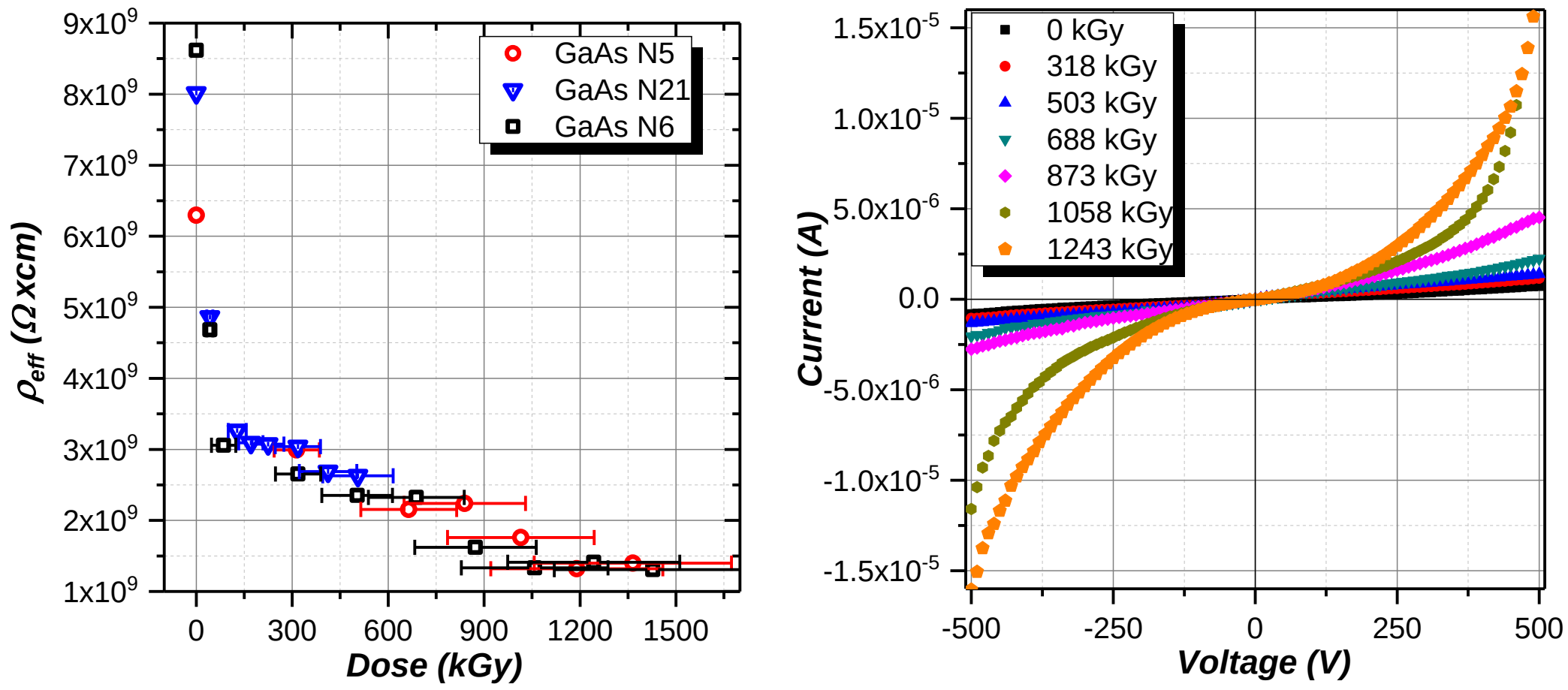

**Figure 6:** (color online) *Dose dependence of effective resistivity for GaAs:Cr N21,5,6 (left) and I-V characteristics for GaAs:Cr N6 (right) before and after irradiation with different doses. Measurement temperature 20°C.*

In accordance with these ideas, in [16] the authors describe the energy levels structure generated when GaAs is irradiated with electrons and gamma rays, indicating the appearance of the shallow E1, E2 and E3 electron traps, located near the middle of the bandgap levels E4 and E5, and the shallow hole traps H1 and H2 near the valence band. It is explained that the studies of defect production by electron irradiation in n-type material provide evidence that the observed traps correspond to primary defects, i.e. vacancies and interstitials in the As sublattice: $V_{As}$ and $As_i$. It is assumed that the E1 - E5 electron traps are formed in the As sublattice are Frenkel pairs ($V_{As}$-$As_i$) [16]. The hole trap H1 refers to arsenic vacancies ($V_{As}$), and H2 – H5 traps to complexes including $As_i$ [16, 21]. It was noted in [16] that due to the low mobility of primary radiation defects in $A^{III}B^{V}$ compounds, the defect spectrum is determined by intrinsic lattice defects, and the contribution of secondary effects — the interaction with defects with impurity atoms is not significant at a temperature of 300 K. We assume that defects formed in HR GaAs: Cr after irradiation by 20.9 MeV electrons are similar to defects formed in GaAs by 1 MeV electrons. In [16] was noted that for GaAs irradiated by 1 MeV electrons, the type and rate of introduction of radiation defects does not depend on the technology of growing the material, nor on the type and concentration of the main alloying substance. These defects have annealing temperature near 500 $^{o}$K [15, 16, 21].

The presence of a slight asymmetry between the I-V and resistivity characteristic, more pronounced in case of GaAs:Cr detector at higher doses, is shown in Figure 6, and this is likely related to the asymmetric potential distribution within the device, which was changed after the radiation damage.

For silicon sensors, an increase in the leakage current at room temperature reaches almost four orders of magnitude for the maximum absorbed dose of 1.5 MGy is shown in Figure 7. In general, the behavior of silicon detectors under irradiation was extensively studied before and the results are reported in, for example [1, 2, 17,19].

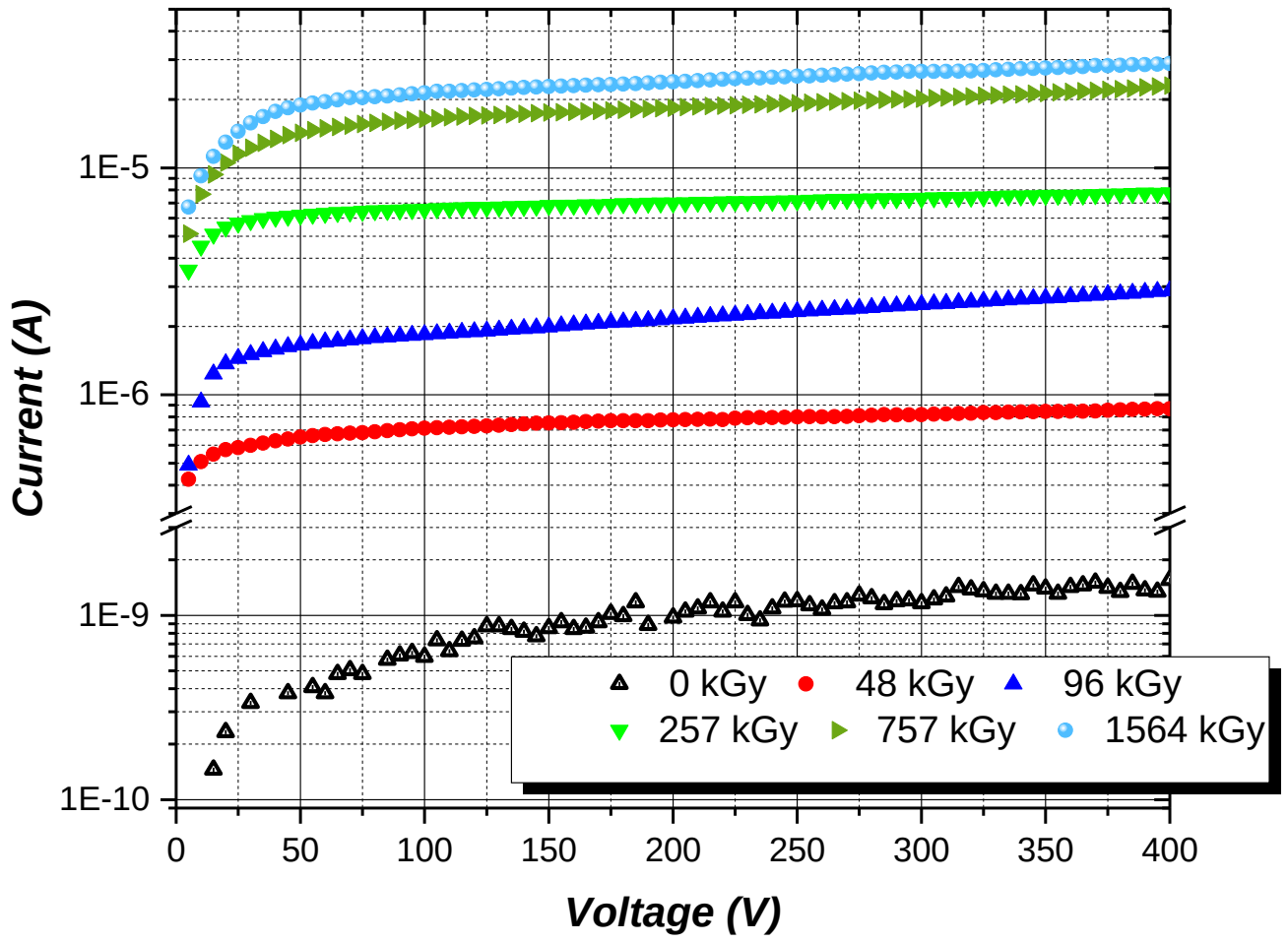

**Figure 7:** (color online) *I-V characteristics for Si N 3 before and after irradiation with different doses. Measurement temperature 20$^{o}$C.*

We can see that the increase in leakage current after irradiation is much more pronounced for Si sensors than for GaAs:Cr at room temperature.

### 5.2. *CCE measurements.*

Figure 8 shows the MIP spectra of Si and GaAs:Cr sensors before irradiation, at room temperature. The pedestal is described by a Gaussian distribution and its width is determined by the readout system resolution. It can be seen for detectors of both types that the pedestal and the MIP signal are well separated. The readout system was calibrated by charge injection from generator and by calibration spectra from γ-sources, the resulting spectra presented on Figures 8, 9, 10, 16, 17 where the charge unit corresponds to 1 ke$^-$.

The MIP spectra for GaAs:Cr sensors after the irradiation doses of 0.5 and 1.5 MGy are presented in Figure 9. Spectra were collected at room temperature (21 °C). We can observe that with an increase in the absorbed dose, the signal peak gradually shifts toward the pedestal and their distributions begin to overlap. At 1.5 MGy the maximum overlap is observed, although the individual signal and pedestal peaks are still distinguishable.

For Si sensors, the MIP spectra after absorbed doses of 0.5 and 1.3 MGy were measured at room temperature and are shown in Figure 10. It is seen that for 0.5 MGy the pedestal and the MIP signal peaks can be identified, and when the dose reaches 1.3 MGy they completely overlap. The signal amplitude and the signal-to-noise ratio decrease as a result of radiation damage. It should be noted that the signal-to-noise ratio decreases both due to a reduction in the signal amplitude and due to the fact that noise becomes more intense as the reverse current grows.

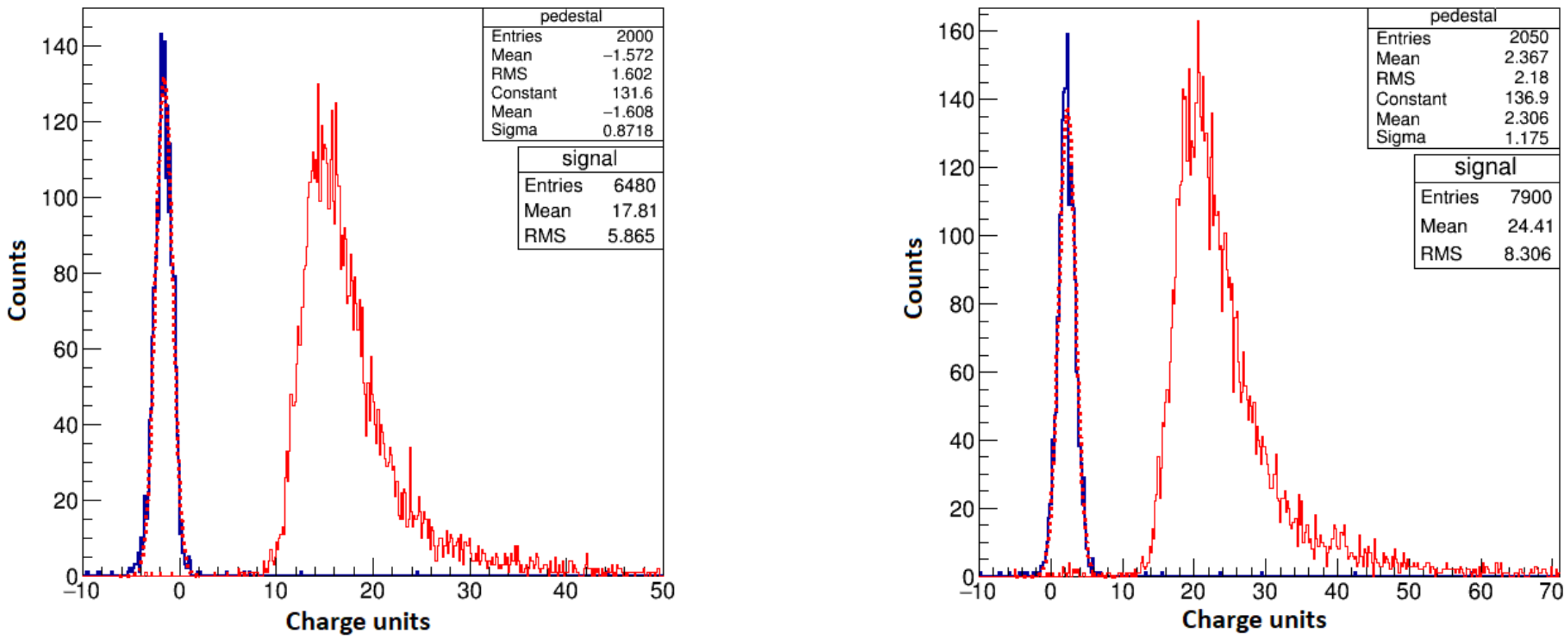

**Figure 8**: *MIP spectra for Si N3(250 μm), $U_{bias}$=100 V (left) and GaAs:Cr N5( 300 μm), $U_{bias}$=-200 V (right) sensors. The pedestal distribution is fit by Gaussian (dash). Measurement temperature 21$^o$C.*

Figure 11 shows the experimental CCE behavior with the irradiation dose normalized to the CCE before irradiation for some selected GaAs:Cr and Si sensors. As stated in [19], after 1 MeV neutron fluence $\Phi_n$~$10^{13}$ cm$^{-2}$, which in our case corresponds to a non-ionizing energy losses (NIEL) for the equivalent 20 MeV electron fluence $\Phi_e$~$2\times10^{14}$ cm$^{-2}$ or a dose of about 50 kGy, the conductivity inversion is observed. This is the transformation of *n*-type bulk detector conductivity into a *p*-type one induced by the formation of radiation defects in silicon. The detector capacitance starts increasing after conductivity inversion and grows further with an increase of dose, since the bulk resistivity continues decreasing. A constant bias of 100 V depletes a thinner and thinner detector region which leads to a decrease in charge collection, especially for thicker Si type 2 sensors, which is good seen in Figure 11.

For GaAs: Cr sensors, the displacement of the MIP signal down to its overlap with the pedestal is due to a decrease in charge collection with the absorbed dose. Such a behavior can be explained by a decrease in the electron lifetime values due to generation of deep levels which acts as traps and recombination centers.

The measurement of pedestal width in a GaAs:Cr shows that it remains almost unchanged during irradiation process. For Si detectors the pedestal considerably broadens due to increased dark current and the measurement becomes difficult for absorbed doses over 0.5 MGy. For both groups of Si sensors the CCE reduction is smaller than for GaAs:Cr detectors. In GaAs:Cr the CCE median falls monotonously and abruptly to the dose of 1 MGy, but then it decreases slowly to the maximum irradiation dose.

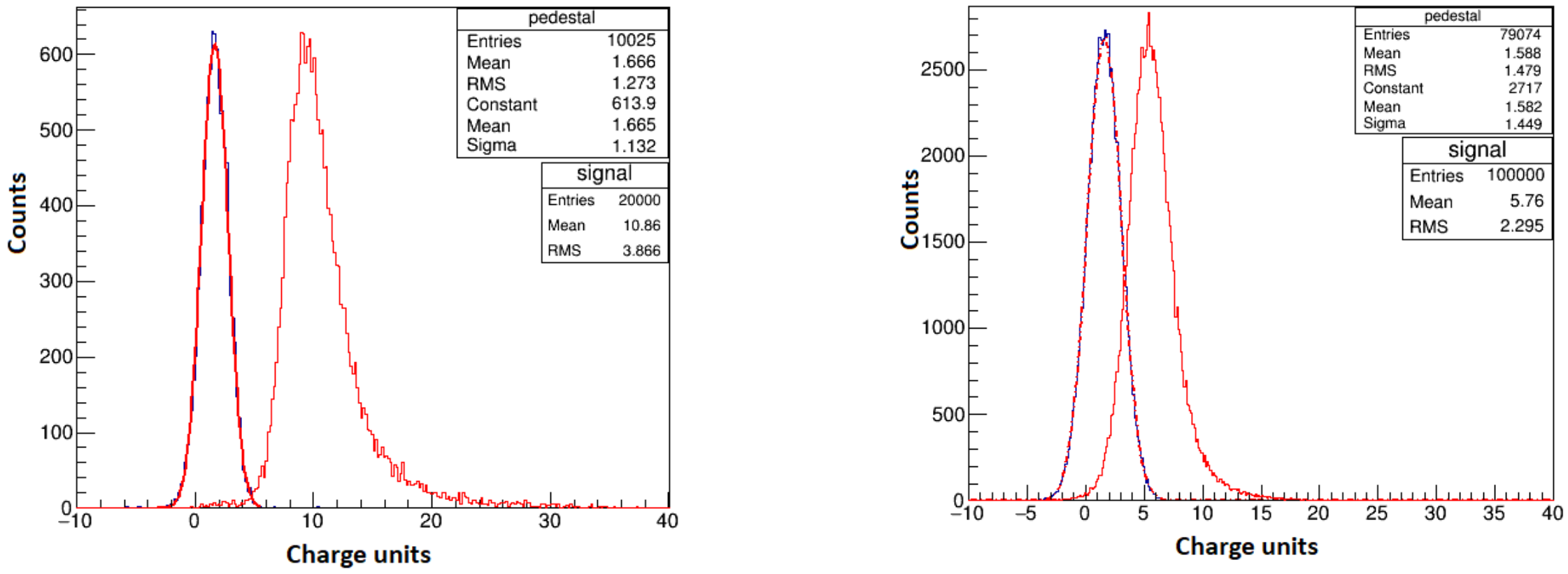

**Figure 9:** *MIP spectra for GaAs:Cr N4 after irradiation dose 0.5 MGy (left) and GaAs:Cr N6 after 1.4 MGy (right).* $U_{bias}$*=-500V. The pedestal distribution is fit by Gaussian (dash). Measurement temperature 21*$^{o}$*C.*

It was observed for all the studied samples. We assume that the concentration of capture centers increases proportionally to the dose, which reduces the electron path. Since the CCE directly depends on the free electron path in the sensor, the fall in the CCE will be inversely proportional to the capture centers concentration and can be represented by the following formula:

$$CCE = \frac{1}{a \times D^b + 1} , \qquad (1)$$

where D is the dose, $\boldsymbol{a}$ and $\boldsymbol{b}$ are normalization factors, and $\boldsymbol{b}$ should be equal to 1 if the drift length of the electrons exceeds the thickness of the sensor ($L_n >> d$). In this case parameter $\boldsymbol{a}$ is proportional to the product of cross section and number of centers. In case of large dose parameter $\boldsymbol{b}$ should be less than 1, because if there are several capture centers on the electron path to the anode, only the first affects to the CCE. As shown in the Figure 11, the experimental CCE on absorbed dose dependence for GaAs:Cr sensors can be very well approximated by the function from the Formula 1.

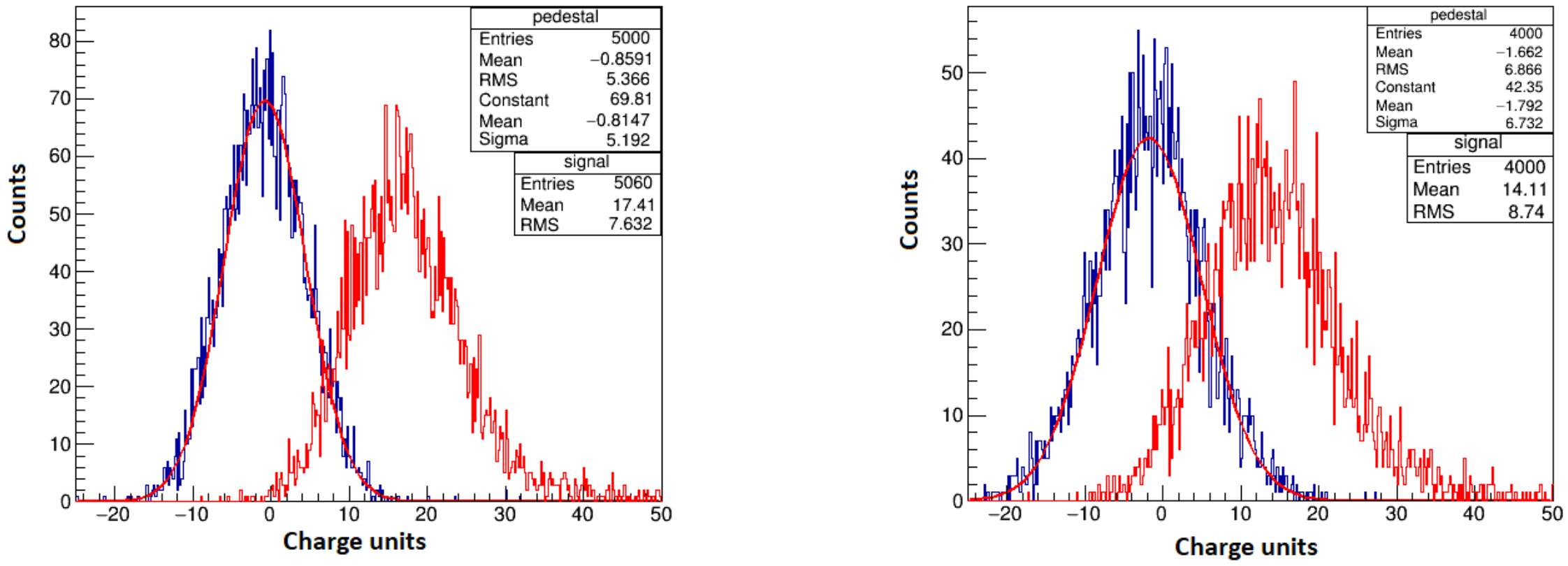

**Figure 10:** *MIP spectra for Si type 1: Si N5 after irradiation dose 0.5 MGy (left) and Si N3 after 1.3 MGy (right). $U_{bias}$=100V. The pedestal distribution is fit by Gaussian (dash). Measurement temperature 21$^{o}$C.*

This fact allows us to assume that CCE deterioration with the dose increase up to 1.4 MGy in the studied GaAs:Cr detectors is due to the creation of new capture centers in the bulk material as result of the 20.9 MeV electron irradiation.

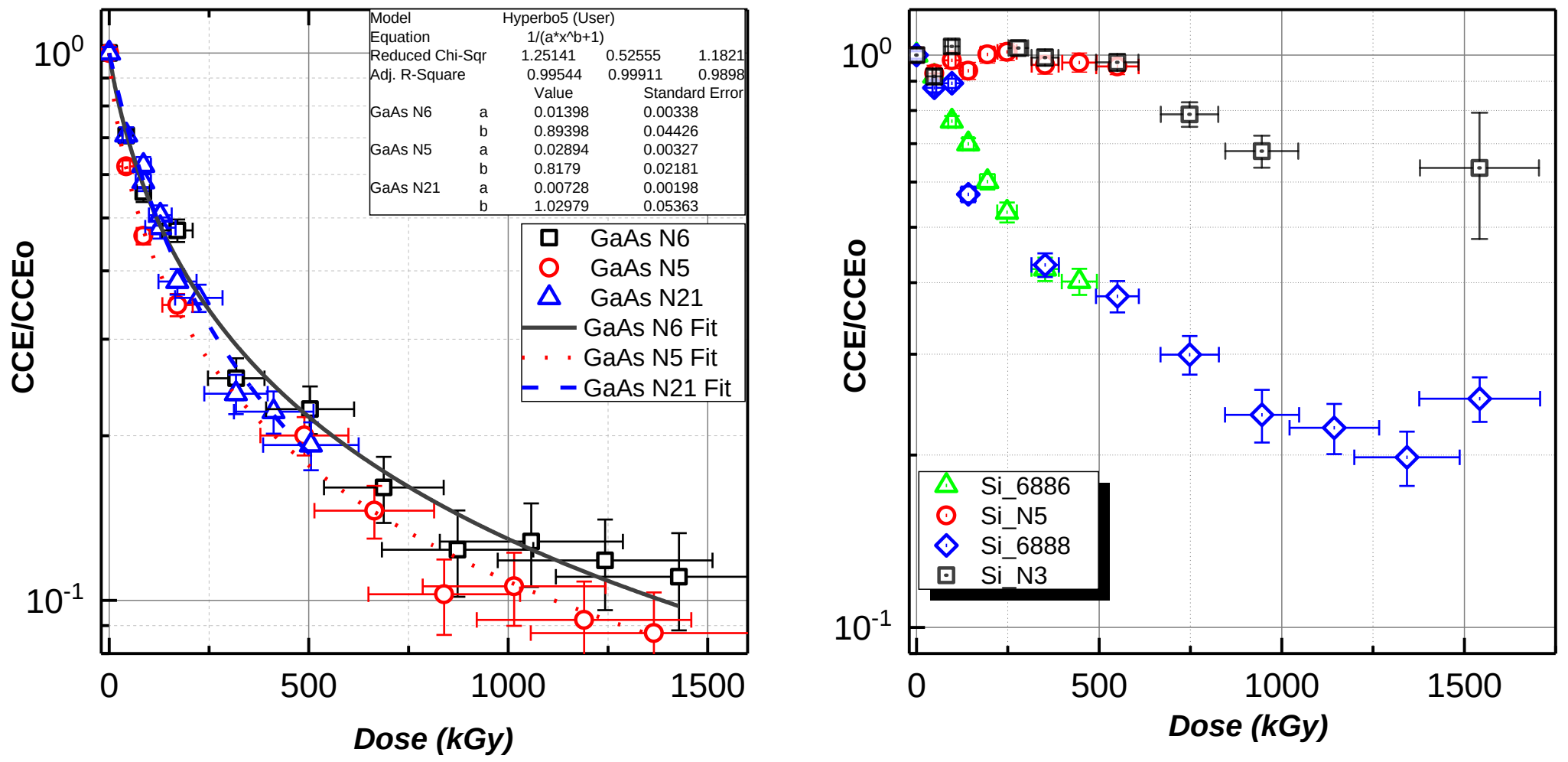

**Figure 11:** *Dependence of CCE on absorbed dose for N5, N6 –(1), N21 – (2) GaAs:Cr sensors, $U_{bias}$=-200 V (left) and Si N3, N5 – type1, 6886,6888 –type2, $U_{bias}$=100V (right). Measurement temperature 21$^{o}$C.*

The CCE behavior as a function of the applied bias voltage for GaAs:Cr is presented in Figure 12. For all unirradiated sensors the CCE reaches the saturation at field strength of about 4 kV/cm. For irradiated sensors, there is a rapid CCE growth up to the field strength of about 8 - 9 kV/cm at first and then a slow growth up to 14 kV/cm followed by saturation. The previously observed CCE deterioration as a consequence of the radiation-induced decrease in charge carriers (electrons for GaAs:Cr) lifetime, can be significantly affected by the electric field strength. At higher field strength the electron path increases and this leads to an increase in charge collection.

The analysis of above presented results leads to the conclusion that with the irradiation the resistivity of the material decreases (the number of free carriers increases), while the charge collection also decreases (the lifetime of the carriers falls);

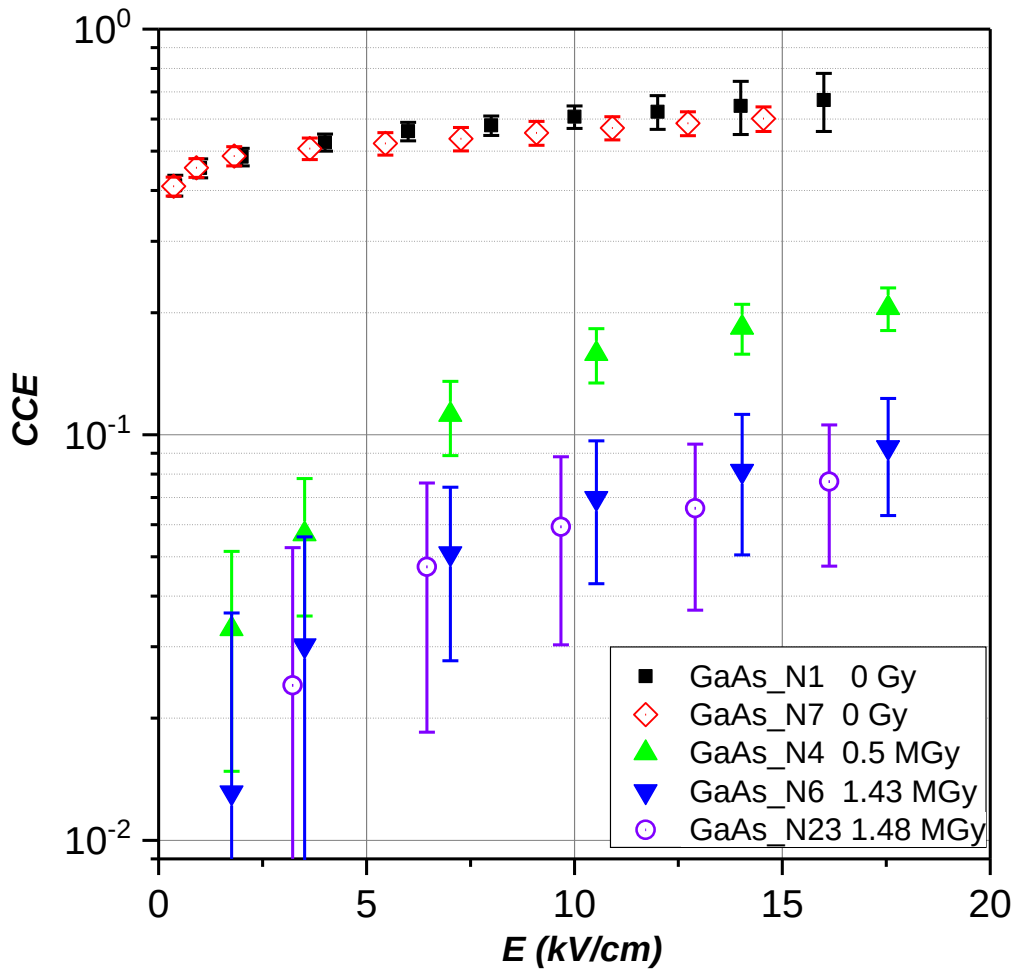

**Figure 12:** *CCE as a function of the electric field strength for unirradiated and irradiated GaAs:Cr sensors at different doses. N1, N7 - unirradiated GaAs:Cr, N4, N6, N23- irradiated GaAs:Cr. Measurement temperature 21°C.*

As can be seen from Figures 8, 9 and 10, the ratio of the width of the pedestal to the average signal is essentially a signal to noise ratio. Thus the width of pedestal determines the resolution of the sensor, i.e. ability to separate a signal from noise. In order to estimate numerically the effect of pedestal broadening, we calculate the K-ratio:

$$\boldsymbol{K} = \frac{\boldsymbol{N}_{\mathbf{2\sigma}}}{\boldsymbol{N}_{\mathbf{total}}}, \qquad (2)$$

where $N_{2\sigma}$ is the number of events when the signal exceeds the pedestal by more than 2σ of pedestal resolution and $N_{\text{total}}$ is the total number of events in MIP spectrum. It is assumed that only MIP signals are represented in the spectrum. The greater this ratio, the better the signal and pedestal are separated in the spectrum. Figure 13 presents the K-ratio dependence on the absorbed dose being studied for all types of sensors. It can be seen that with irradiation for GaAs:Cr sensors K-ratio remains higher than for Si sensors at room temperature.

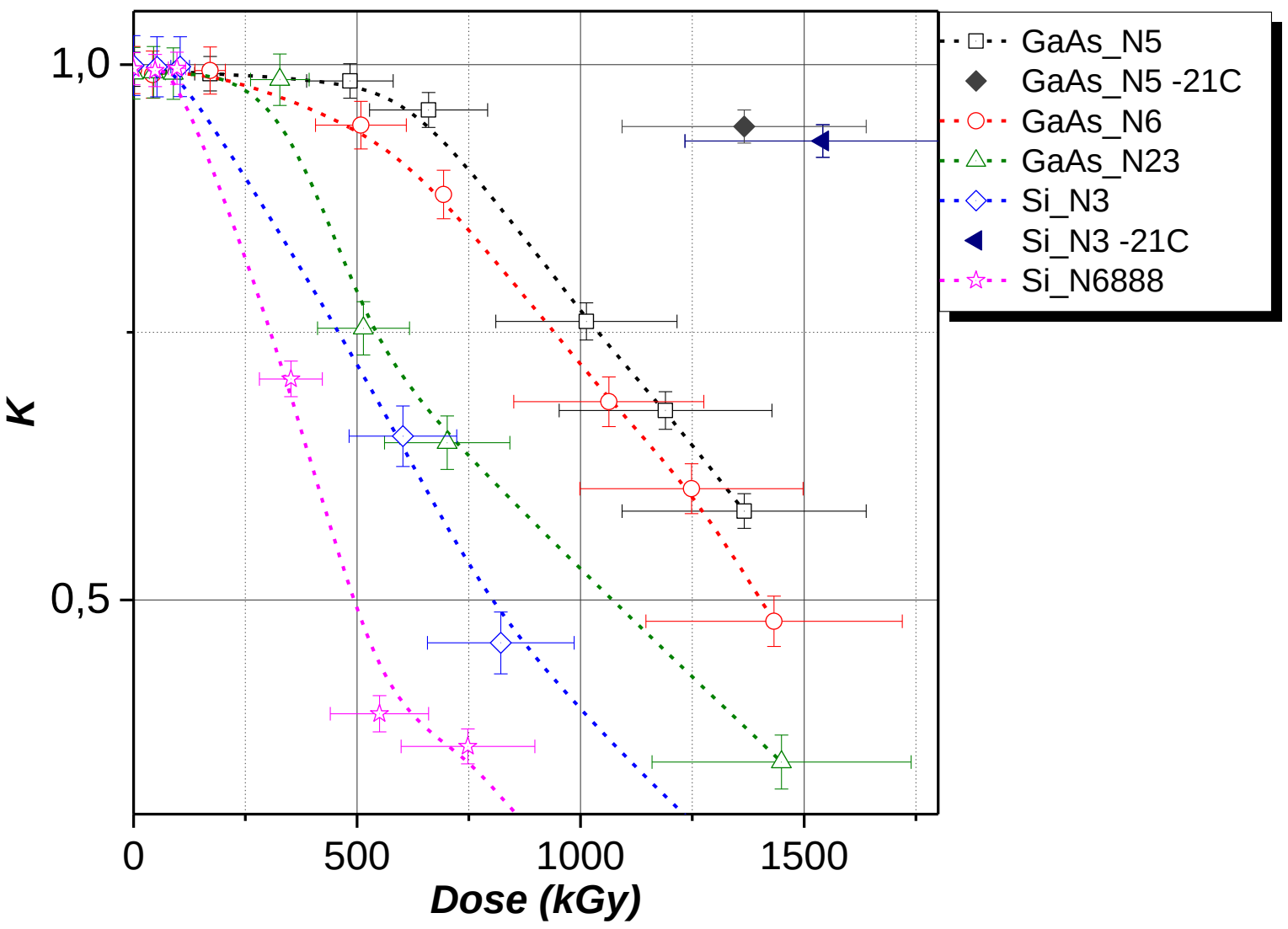

**Figure 13:**(color online) *The K-ratio dependence on the absorbed dose (2σ–criterion), N5, N6 - GaAs:Cr (1), N23 – GaAs:Cr(2), Si_N3 - type1, Si_6888 - type2.* $U_{bias}$ *=100V for Si,* $U_{bias}$ *=-200V for GaAs:Cr, measurement temperature 21°C and -21°C for GaAs:Cr N5, Si N5. The lines are visual guides.*

### 5.3. *I-V and CCE temperature dependences for irradiated sensors.*

For Si and GaAs sensors the I-V dependence on the temperature is typical for semiconductors (Figure 14). The resistivity is given by the following expression:

$$\rho \sim T^{-2} e^{\frac{E}{2kT}} \quad , \qquad (3)$$

where **E** is the energy gap, **T** is the temperature, and **k** is the Boltzmann constant.
As can be seen in Figures 10 and 16, cooling of the Si sensor significantly improves the MIP signal and the pedestal separation in the spectrum. The different CCE dependence between two types of Si on the Figure 11 can be explained by the insufficient applied voltage at different thicknesses of the sensors. On the Figure 11 for Si sensors CCE was measured at $U_{bias}$ = 100V, when Figure 15 shows that after irradiation with a dose of 0.5 MGy for Si type 1 the full depletion voltage is about 150 V when for Si type 2 it is about 300V. Thus, the voltage applied to the Si sensors after irradiation was insufficient to fully collect the charge. At low temperatures it becomes possible to increase the bias voltage for Si sensors and get maximum charge collection (Figures 16, 18). Figure 17 shows that MIP spectra, collected at different temperatures, are similar for GaAs:Cr, although the pedestal width reduces slightly at lower temperatures.

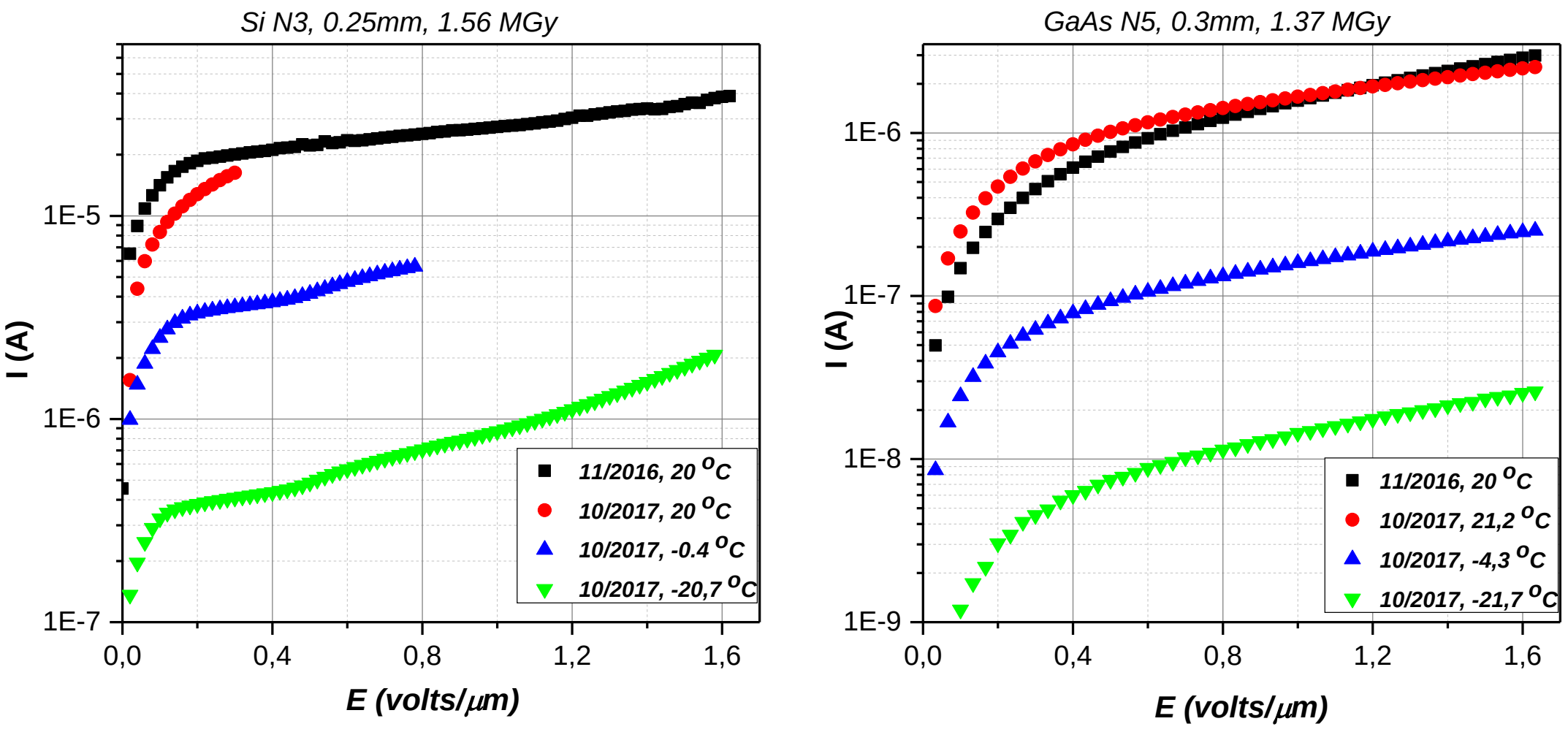

**Figure 14:**(color online) *I-V characteristics measured at different sensors temperatures: Si _N3 (type1) after irradiation dose of 1.56 MGy (left) and GaAs:Cr N5 after irradiation dose of 1.37 MGy (right).*

Figures 18, 19 show that CCE does not depend on the temperature for all types of sensors. For Si (type 2) and GaAs:Cr sensors CCE strongly depend on the applied voltage.

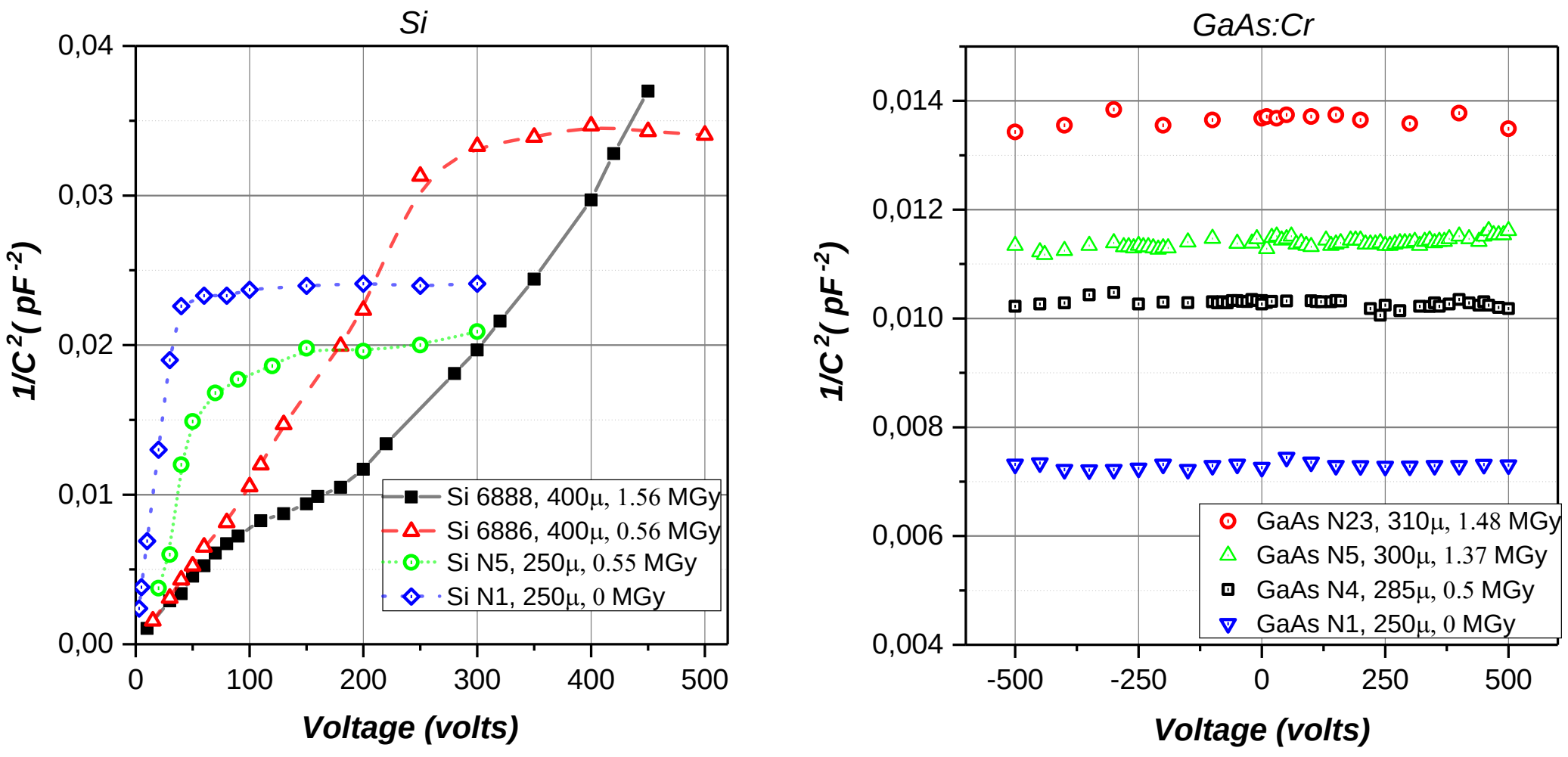

**Figure 15:**(color online) *$C^{-2}$-V for sensors Si type1: N1, N5; Si type2: 6886, 6888 (left) and GaAs:Cr (right) at different doses. Measurement frequency 10kHz, temperature 21°C. The lines are visual guides.*

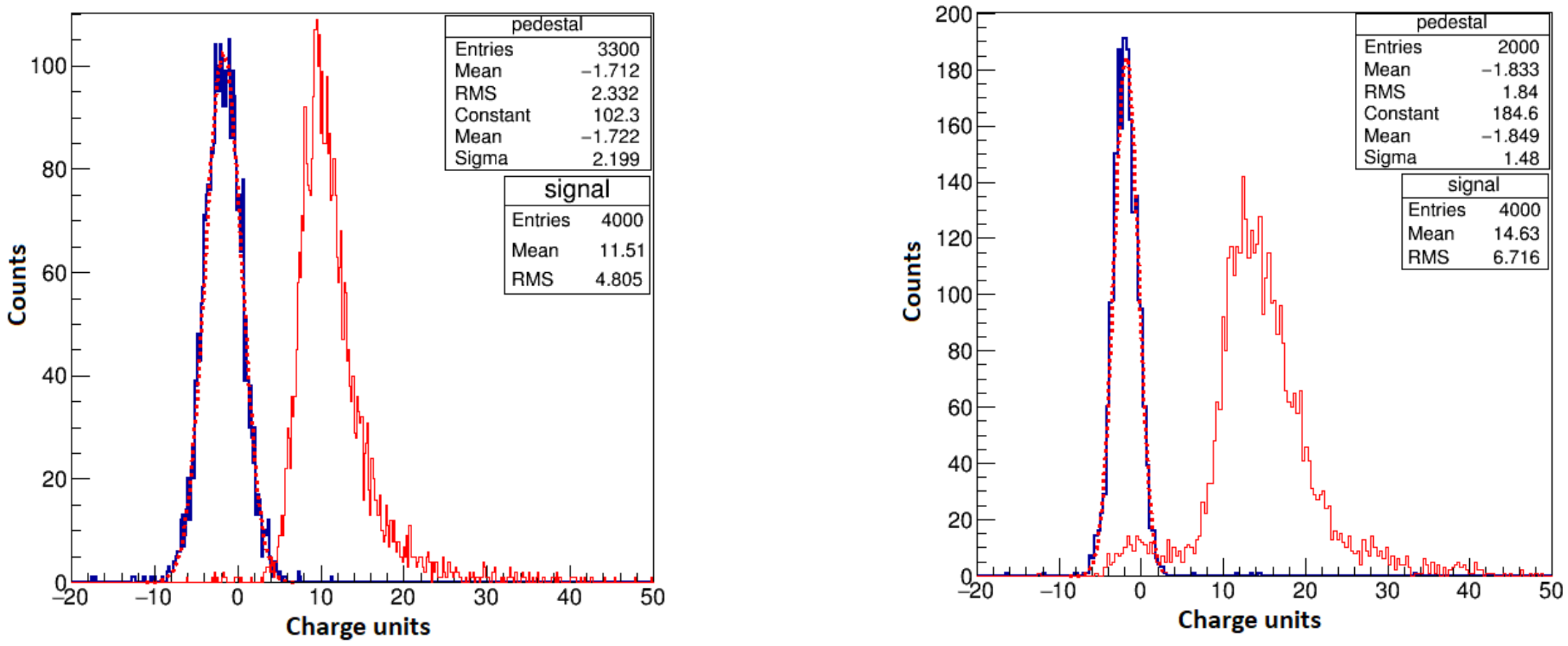

**Figure 16:** *MIP spectra for Si sensors: N3 type1, $U_{bias}$=100V (left) and N6888 type 2, $U_{bias}$=400V (right) after irradiation dose of 1.56 MGy. The pedestal distribution is fit by Gaussian (dash). Measurement temperature -21$^{o}$C.*

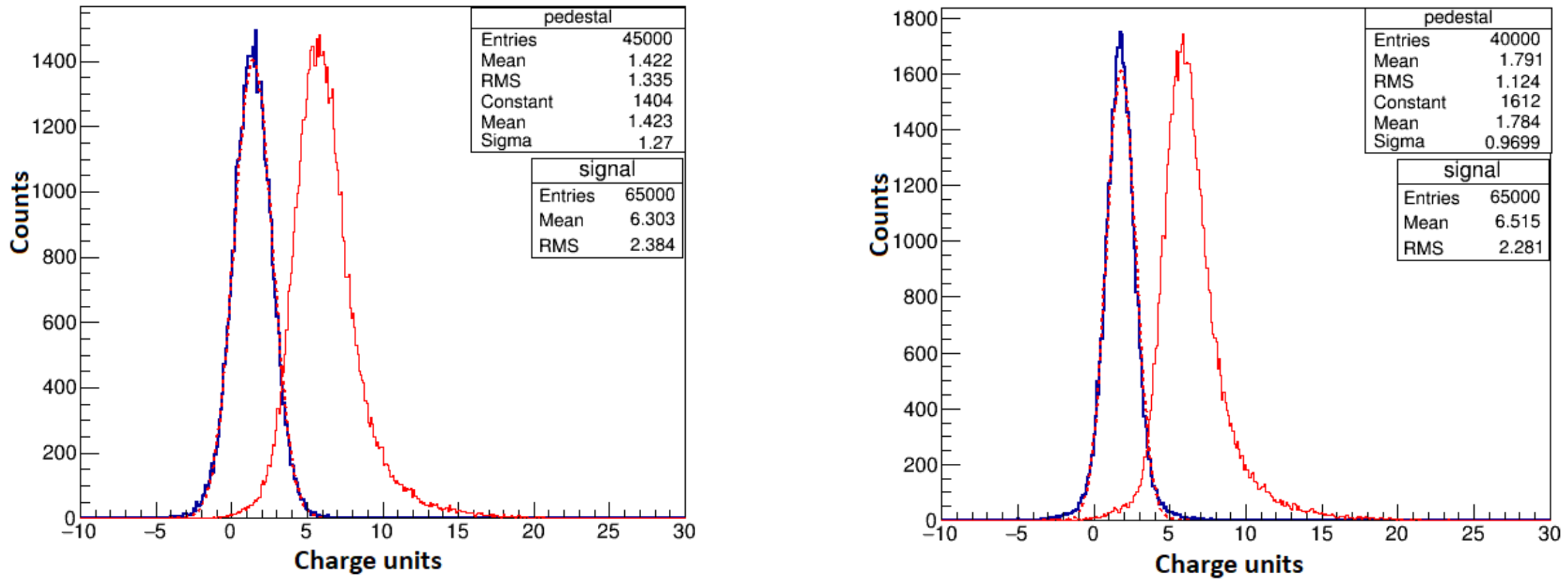

**Figure 17:** *MIP spectra for GaAs:Cr N5 T=20°C (left) and T=-4°C (right), $U_{bias}$=-500V, irradiation dose of 1.37 MGy. The pedestal distribution is fit by Gaussian (dash).*

The equivalent noise charge is proportional to the square root of the product of the dark current to the integration time. For both types of Si sensors the pedestal width determined by the reverse current strongly drops with cooling which is in good agreement with previous studies [17, 18, 19].

As it is seen in Figure 18, for Si type 1 the CCE saturates at the level of 80% with $U_{bias}$>200V, for Si type 2 it does not reach saturation, but becomes more than 50% with $U_{bias}$ = 500V. That correlates with C-V characteristics and means that the bias voltage should be higher than the full depletion voltage in order to get the maximum charge collection.

The Si sensors were stored in the refrigerator, while the GaAs:Cr ones were at room temperature. I-V, CCE and the pedestal width measured just after irradiation in 2016 and one year after show no significant changes (Figures 14, 18, 19). For GaAs:Cr, CCE has no saturation with the applied voltage up to 3V/μm. The temperature dependence of the energy resolution is less significant for GaAs: Cr sensors than for Si ones (Figure 19). However, at room temperature, when $U_{bias}$ exceeds 500V, the intrinsic noise of the sensor increases and becomes comparable to the resolution of the measurement setup (which has a RMS ~ 1ke$^-$) mainly due to an increase in the sensor current.

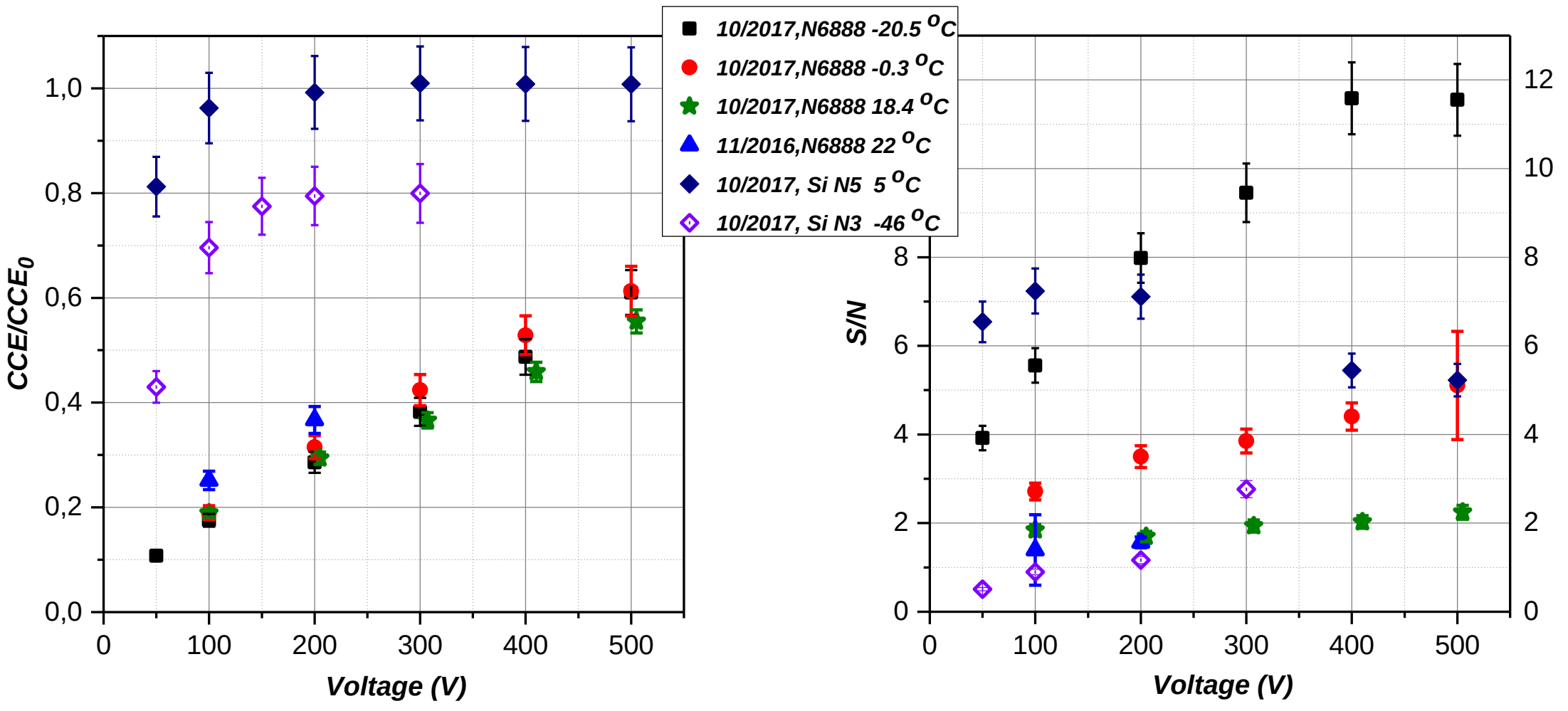

**Figure 18:** *CCE and the equivalent signal to noise ratio(S/N) as a function of applied voltage for Si N3 and N6888 after irradiation dose of 1.56 MGy, Si N5 after irradiation dose of 0.55 MGy, measured with different temperatures.*

Nevertheless, after irradiation GaAs: Cr sensors remain more preferable for some systems where sensor cooling is difficult to implement, although setup configuration and measurement conditions must be taken into account. If cooling is possible, Si sensors are more preferable due to small CCE reduction.

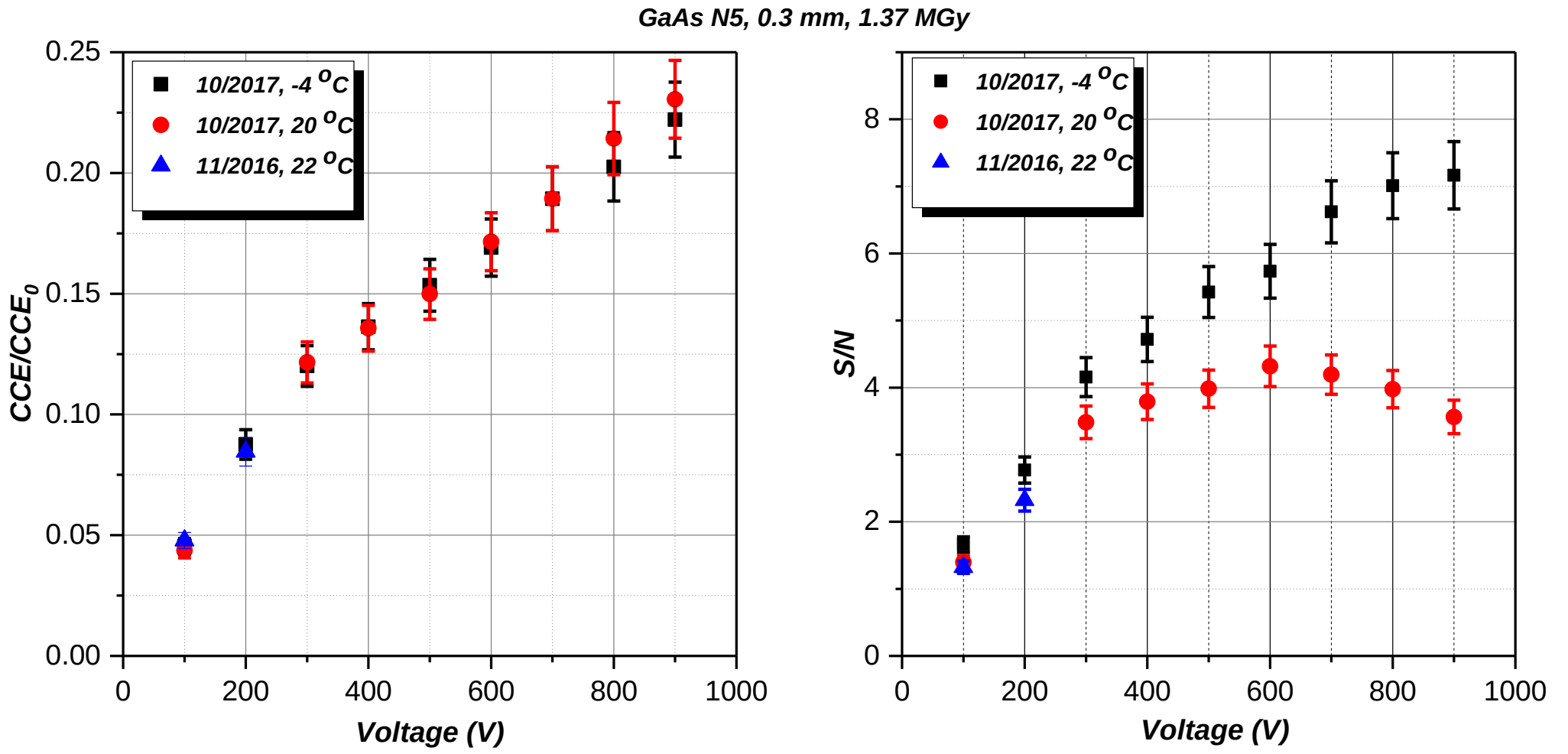

**Figure 19:** *CCE and the equivalent signal to noise ratio(S/N) as function of applied voltage for GaAs:Cr N5 after irradiation dose 1.37 MGy, measured with different temperatures.*

## 6. Conclusion

Radiation damages caused by 20.9 MeV electrons lead to different results in GaAs:Cr and Si sensors. The dark current increased 3-7-fold at the maximum achieved doses of 1.4 MGy in irradiated GaAs:Cr sensors. This is likely due to the introduction of displacement damages in the material, corresponding to primary and complex defects that lead to the formation of new deep levels in the band gap making electrons transition from the valence band to the conduction band easer. According to [15,16,21], radiation defects, such as E4, have an electron capture depth and cross section close to the analogous growth parameters of EL2 defects. This leads to a decrease in the mean free path of charge carriers and leads to a decrease in charge collection. The concentration and activation energy of radiation defects were not measured in this work.

The storage at room temperature during one year did not lead to any significant changes in the properties of the irradiated sensors. According to [16], the annealing temperature of the shallow defect E1 (Ea = 0.045 eV), as well as other deeper defects E2-E5 and H1-H5 in HR GaAs: Cr, is near 500 K, which is much higher than room temperature. This allows to conclude that no significant annealing of radiation induced damage occurs in GaAs:Cr sensors at room temperature.

Measurements of the pedestal spectra showed that the pedestal width in GaAs:Cr remains virtually unchanged for the whole range of absorbed doses. Sensor exposure to a dose of 1.5 MGy reduces CCE to 10% of the initial value. Increasing the detector bias voltage leads to increase in carrier drift velocity and results in a decrease of trapping probability. This effect allows to improve the CCE up to 20% of initial one.

For both GaAs:Cr and Si sensors the CCE does not depend on the temperature but cooling reduces the dark current, which improves the resolution, especially for Si sensors. For Si sensors the charge collection remains above 80% with the dose of 1.5 MGy (which corresponds to a fluence $\Phi_n \sim 4\times10^{14}$ cm$^{-2}$ equivalent NIEL of 1 MeV neutron). C-V characteristic of irradiated sensors demonstrates that the full depletion voltage is rising with irradiation for both types of Si sensors, more strong for HPK sensors due to greater thickness and correlates well with the behavior of CCE. The capacitance of the GaAs:Cr sensors is independent on the voltage and doesn't change with irradiation. It is equal to the flat capacitor with a thickness of the sensor.

Thus, after the irradiation by 20 MeV electrons up to dose 0.5 MGy (which corresponds to a fluence $2.5\text{x}10^{15}$ $e^{-}\times$cm$^{-2}$ or equivalent fluence $1.3\text{x}10^{14}$ n$\times$cm$^{-2}$ of 1 MeV neutron), registration of ionizing particles using GaAs:Cr sensors may be preferable since they do not require cooling. But charge collection in GaAs: Cr decreases faster than in silicon during irradiation, and the use of cooled silicon sensors is preferred for detecting individual particles.

## 7. Acknowledgments

The authors are grateful to the support of the BMBF-JINR program for detector R&D. We would like to express thanks our colleagues B. Schumm and V. Fadeyev from SCIPP UCSC for providing Hamamatsu Photonics n-type Si sensors, as well as D. L. Demin,

A.P.Sumbaev and the LINAC-200 accelerator group for the helpful and useful discussion during this work.